\documentclass[%
 aip,
superscriptaddress,
amsmath,amssymb,
preprint,%
]{revtex4-1}

\usepackage{graphicx}
\usepackage{dcolumn}
\usepackage{bm}

\usepackage[utf8]{inputenc}
\usepackage[T1]{fontenc}
\usepackage{mathptmx}
\usepackage{xcolor}
\usepackage{soul}
\begin{document}

\preprint{AIP/123-QED}

\title[]{Efficient {\it ab initio} calculation of electronic stopping in disordered systems via geometry pre-sampling: application to liquid water}

\author{Bin Gu}
\affiliation{Department of Physics, Nanjing University of Information Science and Technology, Nanjing 210044, China}
\affiliation{Atomistic Simulation Centre, Queen's University Belfast, Belfast BT71NN, Northern Ireland, United Kingdom}
\author{Brian Cunningham}
\affiliation{Atomistic Simulation Centre, Queen's University Belfast, Belfast BT71NN, Northern Ireland, United Kingdom}
\affiliation{Centre for Theoretical Atomic, Molecular and Optical Physics, Queen's University Belfast, Belfast BT71NN, Northern Ireland, United Kingdom}
\author{Daniel Mu\~noz Santiburcio}
\affiliation{CIC Nanogune BRTA, Tolosa Hiribidea 76, 20018 San Sebastian, Spain}
\author{Fabiana Da Pieve}
\affiliation{Royal Belgian Institute for Space Aeronomy, Av Circulaire 3, 1180 Brussels,
Belgium}
\author{Emilio Artacho}
\affiliation{CIC Nanogune BRTA, Tolosa Hiribidea 76, 20018 San Sebastian, Spain}
\affiliation{Donostia International Physics Center (DIPC), Tolosa Hiribidea 76, 20018 San Sebastian, Spain}
\affiliation{Ikerbasque, Basque Foundation for Science, 48011 Bilbao, Spain}
\affiliation{Theory of Condensed Matter, Cavendish Laboratory, University of Cambridge, Cambridge CB3 0HE, United Kingdom}
\author{Jorge Kohanoff}
\email{j.kohanoff@qub.ac.uk}
\affiliation{Atomistic Simulation Centre, Queen's University Belfast, Belfast BT71NN, Northern Ireland, United Kingdom}
\date{\today}

\begin{abstract}

Knowledge of the electronic stopping curve for swift ions, $S_e(v)$, particularly  around the Bragg peak, is important for understanding radiation damage. Experimentally, however, the determination of such feature for light ions is very challenging, especially in disordered systems such as liquid water and biological tissue.  Recent developments in real-time time-dependent density functional theory (rt-TDDFT) have enabled the calculation of $S_e(v)$ along nm-sized trajectories. However, it is still a challenge to obtain a meaningful statistically averaged $S_e(v)$ that can be compared to observations. In this work, taking advantage of the correlation between the local electronic structure probed by the projectile and the distance from the projectile to the atoms in the target, we devise a trajectory pre-sampling scheme to  select, geometrically, a small set of short trajectories to accelerate the convergence of the averaged $S_e(v)$ computed via rt-TDDFT. For protons in liquid water, we first calculate the reference probability distribution function (PDF) for the distance from the proton to the closest oxygen atom, $\phi_R(r_{p{\rightarrow}O})$, for a trajectory of a length similar to those sampled experimentally. Then, short trajectories are sequentially selected so that the accumulated PDF reproduces $\phi_R(r_{p{\rightarrow}O})$ to increasingly high accuracy. Using these pre-sampled trajectories, we demonstrate that the averaged $S_e(v_p)$ converges in the whole velocity range with less than eight trajectories, while other averaging methods using randomly and uniformly distributed trajectories require approximately ten times the computational effort. This allows us to compare the $S_e(v_p)$ curve to experimental data, and assess widely used empirical tables based on Bragg’s rule.

\end{abstract}

\maketitle
\section{\label{sec:1}INTRODUCTION}

When an energetic ion travels through matter it slows down by continually transferring energy to the electronic and nuclear degrees of freedom of the target\cite{Bragg1905,Bohr1913}. The total energy transfer rate of these interactions is called stopping power ($S$) and it is generally divided into electronic ($S_e$) and nuclear ($S_n$) components. The stopping power is defined as the (kinetic) energy lost by the ion per unit distance travelled: 
\begin{equation}
S=-dE_k/dl=S_e+S_n, 
\label{eq1}
\end{equation}
in which $E_k$ is the kinetic energy of the ion and $l$ is the distance travelled in the target. The word {\it power}, which for historical reasons is used in the sense of {\it time rate}, can be misleading as, strictly speaking, $S$ is equal to the total retarding {\it force} acting on the particle. $S$ is a function of the ion's velocity ($v_{p}$) and the physical and chemical properties of the ion and the target material. 

The $S_e(v_p)$ curve is of great importance for most applications related to ion irradiation of matter including, but not limited to, radiation medical (cancer) therapies \cite{KRAFT2000,Baskar2012,Solovyov2016}, modification of materials \cite{Calcagno1992,Was2007}, next generation power technology \cite{Granberg2016}, and space radiation-induced effects in astronauts \cite{Cucinotta2006,ferrari2009cosmic,cucinotta2012}, spacecraft components and on-board equipment \cite{Jiggens2014,Duzellier2005}.
For example, the great advantage of ion-beam cancer therapy is that it is able to deliver a significant amount of energy to the region where the tumor is located while minimizing the deposition in surrounding healthy tissue. This is achieved by utilizing the fact that the $S(v_{p})$ curve exhibits a pronounced maximum at a specific ion velocity. By choosing the initial energy of the projectile one can tune the distance travelled by the ion through the target until its velocity becomes similar to that of the electrons in the various electronic shells. At that point the ion deposits most of its remaining energy in a very short distance, thus leading to a sharp maximum in the dose vs depth curve, called the {\it Bragg peak}\cite{Race_2010}. Since the velocity of electrons in the valence orbitals of the target atoms depends only weakly on the material, the position of the peak in $S_e(v_p)$ is largely independent on the target and, when portrayed as a function of the projectile's energy $E_k$, it scales with the mass of the projectile and lies typically in the region of 10 keV/nucleon \cite{ziegler2008srim}. At Bragg peak velocities the ion moves so fast that there is no time for the nuclei in the target to react and, hence, nuclear stopping $S_n$ is negligible, while $S_e$ is dominant. This regime is inherently non-adiabatic; the electrons follow their own dynamics while the nuclei stay still.

Since the landmark experiments of Bragg \cite{Bragg1905} and Rutherford \cite{Rutherford1911}, the scientific literature accumulated a large amount of experimentally determined stopping power data and ion range distributions for ion beams with energies above 1 MeV/nucleon \cite{Rutherford1911,Blanchin1969,Ziegler2010}. Experimental measurements of electronic stopping power for ions with energy below a few hundred keV/nucleon in amorphous and liquid targets are not trivial \cite{Shimizu2009,Shimizu2010,Siiskonen2011,Garcia-Molina2013}.
Despite the fact that proton beams have been widely used for cancer therapy, the stopping curve for protons in liquid water has not been determined experimentally in the region of the Bragg peak and below ($E_k<100$ keV). This is significant, because liquid water is overwhelmingly used to model the average biological medium \cite{Yao}. Experimental data are available only at proton energies above 0.3 MeV \cite{Shimizu2010,Yao}. Therefore, any information below this energy is based on extrapolation from higher-energy experimental data, using theoretical models and Bragg's rule with empirical scaling (see below). A better knowledge of the mechanisms through which ionizing radiations interact with water is still of great importance, for both enhancing the effectiveness of cancer treatments and to better develop radiation protection measures for medical and nuclear facilities and the radiation-induced effects in Space missions, when protons are slowed down by or produced as secondary particles through shielding materials, in spacecraft or in planetary subsurface habitats.

Since the early times of quantum mechanics, empirical models of (electronic) stopping power were proposed following different approximations. These include the Bethe-Bloch formalism for high velocity \cite{Sigmund2014}, Fermi-Teller formalism in the low velocity limit \cite{Fermi1947}, and Lindhard's linear response theory (LRT) \cite{Lindhard1964}. The combination of LRT with {\it ab initio} electronic structure calculations improved significantly the quality of electronic stopping results \cite{Campillo1998}. More recently, straightforward real-time electron dynamics simulations allowed for accessing the whole energy spectrum \cite{pruneda2007,Sigmund2014,Correa2018} 
beyond the region of validity of LRT, which deteriorates for projectile energies below 
$\sim 200$ keV/nucleon.\cite{Yost2016SiC} 
By combining a variety of semi-empirical models based on LRT, Monte Carlo track-structure (MCTS) codes, such as Geant4-DNA\cite{Geant4DNA2018}, KURBUC\cite{Liamsuwan2011} or PARTRAC\cite{Alcocer-Avila2019}, can be used to estimate microdosimetric parameters by simulating the slowing-down of particles in 
water \cite{Nikjoo2016,Dingfelder2000,Emfietzoglou2003,Abril2011,DeVera2015}, using an energy loss function that is extracted from experimental data and interpolated for several energies to cover the paucity of measurements. However, the semi-empirical models can only be used under some specific and ideal conditions. There are serious ambiguities when these formulas are used to model a real material. 
For example, during the irradiation process, the charge state of the projectile changes continuously, while the local electronic structure of the target along the path of the projectile also evolves in time \cite{ChristopherRace2011,Correa2018,Sand2019}. The band structure of the target \cite{Lin2008,Ullah2015} as well as electron-phonon coupling at low velocities \cite{Rizzi16} can also contribute to the stopping of ions.  With all these entangled aspects, it is quite difficult to interpret the stopping power for different experiments in various regimes through these analytic models \cite{Correa2018}. 

The most generally accepted and widely used reference data for stopping power comes from SRIM \cite{Ziegler2010} and PSTAR (for proton) \cite{pstar} tables. These have been derived by empirically scaling the limited available experimental data according to Bragg's rule or via Bethe-Bloch formalism. Bragg's rule estimates the stopping power of a compound by adding stoichiometrically the stopping powers of its constituent elements, which are relatively easier to measure than those of compounds. \cite{Braggrule1983} Since this approach generally overestimates experimental values measured in compounds, a scaling factor is applied to the Bragg curve, so that the scaled curve coincides with experimental data points. For protons in liquid water, the scaling factors used by SRIM is 0.94 \cite{ziegler2008srim, Ziegler2010}. These empirical scaling factors originate from the chemical nature of the compound. They are derived by simultaneously fitting the stopping power in several compounds for some given ion velocities, thus yielding the so-called {\it bonding corrections} \cite{CABweb}. The scaling is uniformly applied to the entire stopping curve, without introducing any dependence on the ion's velocity. To the best of our knowledge, there is no independent verification, either theoretical or experimental, of the universal validity of this scaling scheme. 

Recent developments in real-time simulations within time-dependent density functional theory (rt-TDDFT) made it possible to calculate electronic stopping power $S_e$ for high-energy ions travelling across a target material, following specific trajectories \cite{ChristopherRace2011,Correa2018,Correa2012,Schleife2015,Yost2017,Ullah2018,Maliyov2018,LI201841,PhysRevA.100.052707}. However, even with the most powerful high performance computers, rt-TDDFT simulations can only access nanometer-sized model targets. Actual samples used in experimental setups are in the order of micrometers or thicker. Therefore, rt-TDDFT simulations sample only a very short segment of experimentally realized trajectories.
Since $S_e(v_p)$ is sensitive to the local electronic structure of the target along the ion trajectory, an accurate comparison of rt-TDDFT calculations with observations requires a statistical reconstruction of the long experimental trajectory using a large number of nanometer-scale rt-TDDFT simulations. \cite{Rutherford1911,Reeves2016,Schleife2015,Yao}. Yao {\it et al} demonstrated that, for protons in water, the ensemble average of $S_e(v_p=8.0~\mathrm{a.u.})$ based on 64 regularly distributed trajectories in a 16 \AA{}-long water box converges very well to SRIM's data \cite{Yao}. This value $v_p=8$ a.u. is a rather high velocity, corresponding to a proton kinetic energy of 1.6 MeV/nucleon. At such a high velocity, the variation of the stopping power for different ion trajectories is relatively small when compared to the variation observed around the Bragg peak, which is located at $v_p\approx 1.7$ a.u.

In principle, rt-TDDFT calculations can be used as "virtual" experiments to determine $S_e(v_p)$ in different kinds of materials and for various ionic projectiles. Such simulations carry significant advantages over both empirical models and real experiments. However, the huge computational cost required to achieve accurate ensemble averages is a major limiting factor that precludes the general, systematic use of this technique. Therefore, an efficient, low-cost, smart sampling scheme for nanometer-sized trajectories is required, so that the local electronic structure is accurately portrayed. Only after this has been achieved, rt-TDDFT calculations can be extensively used for ion radiation engineering applications. 

To bridge the gap between the experimentally observed $S_e(v_p)$ and the one calculated with a limited set of rt-TDDFT simulations, we will exploit the correlation between the local electronic structure sensed by the swift ion and the distance from the ion to the nuclei in the target sample. To this end,  we propose here a trajectory pre-sampling method that uses only geometric information about the target sample to select a small number of representative short trajectories for the rt-TDDFT simulations, thus reducing significantly the computational cost. The method is based on computing the normalized probability distribution function (PDF) of the distance between the projectile and its nearest nuclei in the target, $\phi(r_{m{\rightarrow}X})$, and comparing it to the converged PDF for a long trajectory, which we call {\it reference} PDF, $\phi_R(r_{m{\rightarrow}X})$. Here $X$ indicates the atomic species in the target, i.e. O and H for water. We then sort the short trajectories according to how close they match the reference distribution, $\phi_R(r_{m{\rightarrow}X})$, and reconstruct the latter by considering trajectories in order of decreasing score.

This scheme is demonstrated by calculating electronic stopping for protons in liquid water. Liquid water is chosen as an example for two reasons. Firstly, as the most essential component of biological tissue, it is regularly used as a model for radiation dose estimations in medical physics \cite{Solovyov2016}. Secondly, liquid water is representative of the more general class of disordered materials. As to proton beams, they are widely used in radiation applications, such as cancer therapy and materials processing \cite{Solovyov2016,Calcagno1992}. In addition, energetic protons are the most abundant species present in solar energetic particles and galactic cosmic radiation, thus posing a threat to astronauts and on-board instruments in space missions \cite{cucinotta2012,slaba2017optimal,Rostel2020} by both  the primary and secondary ionizing radiations. The calculation of $S_e$ for protons in liquid water has great value both theoretically and in terms of applications. 

The paper is organised as follows. Theory and methods used for the {\it ab initio} simulation of $S_e$ are described in Section~\ref{sec2}, together with the  trajectory pre-sampling scheme proposed here. Results and discussions are presented in Section ~\ref{sec3}, while in the last Section we summarize the conclusions and offer suggestions for future studies and for other possible applications of the scheme.  

\section{\label{sec2} Theory and Methods} 
\subsection{\label{sec2.1}rt-TDDFT calculation of $S_e(v_p)$ along single ion trajectory}
There are two main contributions to electronic stopping when a swift ion moves through a target sample: a continuous energy transfer at a practically constant rate, and a superimposed sequence of more violent, step-wise transfers every time the projectile passes close enough to an atom in the target. At high velocities one can describe electronic stopping via linear response approaches, as the perturbation introduced by the projectile is relatively small. However, at lower velocities perturbative methods are not suitable. The most accurate theoretical approach to describe non-equilibrium electron dynamics under ion irradiation is to solve the time-dependent many-body Schr\"{o}dinger equation (TD-MBSE).  However, the direct solution of TD-MBSE is computationally too expensive for the typical system sizes required \cite{Attaccalite2011,Sangalli2019}. 
Density functional theory (DFT) recasts the many-body problem in terms of an auxiliary system of non-interacting electrons via the Kohn-Sham formalism.
\cite{Hohenberg1964,ks1965,kohanoff2006}. Ground state DFT and its time-dependent extension time-dependent DFT (TDDFT)  \cite{Runge84} proved very successful when comparing predictions with experimental observations \cite{Burke2012}. In TDDFT, the one-electron Kohn-Sham orbitals evolve according to the time-dependent Kohn-Sham equations. TDDFT, its applications, and numerical implementations have been recently reviewed \cite{Ullrich2011,Maitra2016}.  In the seminal work of Pruneda et al. in 2007, TDDFT was used to compute electronic stopping power in materials by simulating a swift ion travelling through a target material \cite{pruneda2007,Correa2018}. 

When a projectile is forced to move at a constant speed through a target material, the total energy of the system will increase by an amount $\Delta{E}$ as a result of the work done by the constrain to maintain the projectile's velocity constant \cite{Schleife2012}. For projectile's kinetic energies large enough, i.e. above a few keV/nucleon, the motion of the host nuclei is negligible in the time scale of the projectile's transit, as the response time of the nuclei is much longer. 
Therefore, there is no appreciable effect on the electronic dynamics if the host nuclei are constrained to stay at their initial positions during TDDFT stopping simulations. The advantage is that, in this way, the change in total energy is due only to the electronic subsystem, $\Delta{E_e}$. The electronic stopping power for a projectile's trajectory of length $\Delta{L}$ can then be calculated as: 
\begin{equation}
\begin{aligned}
S_e(v_{p})=\Delta{E_e}/\Delta{L}.
\end{aligned}
\label{eq2}
\end{equation}

Several different mathematical descriptions have been proposed the propagation of Kohn-Sham orbitals within TDDFT 
\cite{Castro2004,Andermatt2016,Correa2018,Maliyov2018}. In this work, we use the rt-TDDFT implementation in CP2K \cite{Hutter2014,Andermatt2016}. Much in the spirit of the SIESTA implementation of rt-TDDFT \cite{Tsolakidis2002,Emilio2017}, in CP2K the Kohn-Sham orbitals are expanded in a  local basis, which in this case consists of Gaussian functions, while the electronic density is represented either in plane waves (GPW) for calculations with pseudo-potentials, or in plane waves augmented with Gaussian functions in the vicinity of the nuclei (GAPW) for all-electron calculations. While a full plane wave implementation as in Qb@ll \cite{Schleife2012,Draeger2017} would have a more controllable basis convergence,
the need for a large cutoff makes it, at present, computationally too expensive for general applications. 
Local-orbital implementations, either numerical as in SIESTA \cite{Soler2002}, or Gaussian basis functions as in CP2K and other codes \cite{Bruneval2016,Maliyov2018}, are significantly more efficient provided a smart choice of basis set is made that captures the possibility of electrons being excited to high energy states or even ejected. Moreover, it has recently been shown that both valence and core electrons can contribute significantly to electronic stopping \cite{Correa2018,Ullah2018,Yao}. Therefore, for the calculation of $S_e$ of protons in liquid water we used the GAPW all-electron implementation of rt-TDDFT in CP2K, together with the (triple-zeta + double polarization) 6-311G** basis set \cite{Krishnan1980,kohanoff2006}, which includes both valence and core electrons. The electronic density was expanded in plane waves up to an energy cutoff of 500 Ry. The electronic Brillouin zone was sampled at the $\Gamma$-point.  
We adopted the adiabatic Perdew–Burke–Ernzerhof (APBE) exchange-correlation functional \cite{Krishnan1980}, which excludes memory effects. 

The first step in the calculation of $S_e(v_p)$ by rt-TDDFT is to set up the initial conditions of the target material. In this work, two independent configurations of liquid water were selected from a Born-Oppenheimer {\it ab initio} molecular dynamics simulation carried out at the same level of accuracy of the ensuing rt-TDDFT simulations. A planar projection of the first configuration (Conf-1) is shown in Fig.\ref{f1}(A). The bulk liquid water sample, consisting of 203 water molecules, is contained in a box of dimensions: 24 \AA{} in the $x$ direction and 16 \AA{} in both the $y$ and $z$ directions. The simulation box is subject to periodic boundary conditions (PBC). The density is 0.996 g/cm$^3$, corresponds to the experimental value at room temperature (300 K) and pressure (1 atm). Unless explicitly stated, all the calculations in this work were carried out based on Conf-1. The second configuration (Conf-2) was used mainly to demonstrate the independence of the geometry per-sampling scheme from the reference configuration of the target material.
The coordinates of Conf-1 and Conf-2 are given in Table S1 of the supplementary material.

To initiate the irradiation simulations we first placed a proton at the initial position of its trajectory. The additional charge was compensated with a uniform negative background. We then re-calculated the ground state electronic structure of the whole simulation system. Next, we started the real-time propagation of the Kohn-Sham orbitals by assigning an initial velocity to the proton. The velocity of the projectile and the configuration of the water sample were maintained unaltered by setting to zero the forces on all nuclei, so that the trajectory of the projectile was rectilinear. The time step $\Delta t$ of the real-time propagation was determined by setting a constant displacement of $\Delta x=0.01$ \AA{} in each integration step, i.e. $\Delta t=\Delta x/v$.  The largest time step used for our simulations is 1.83 attoseconds, corresponding to the smallest velocity of 0.25 a.u. The suitability of this choice was confirmed by comparing the energy transfer profiles with simulations using smaller $\Delta x$. The energy transfer to the electronic sub-system during the rt-TDDFT simulation was monitored and used to compute $S_e$ along individual trajectories according to equation (\ref{eq2}). While simulations were run under PBC, stopping was computed using trajectories that were fully contained within a single simulation box. This was to prevent the projectile from sampling regions that were already electronically excited due to the previous passage of the proton.



\subsection{Geometry dependence of the electronic stopping power}
Since the electronic stopping power depends on the electronic density, it will depend, sometimes quite significantly, on the specific trajectory \cite{Dorado1993,pruneda2007,Sigmund2014,Correa2018}. This is particularly important for nanometer-sized trajectories that sample the target inadequately. Moreover, for such short trajectories the magnitude of the fluctuations in $S_e(v_p)$ from one trajectory to another depends significantly on the projectile's velocity, particularly in the region of the Bragg peak. Achieving efficiently a high accuracy at the Bragg peak is of paramount importance, as this guarantees that results at any other velocity will be well converged.
 
\begin{figure}[htbp]
\centering
\noindent\includegraphics[width=12cm]{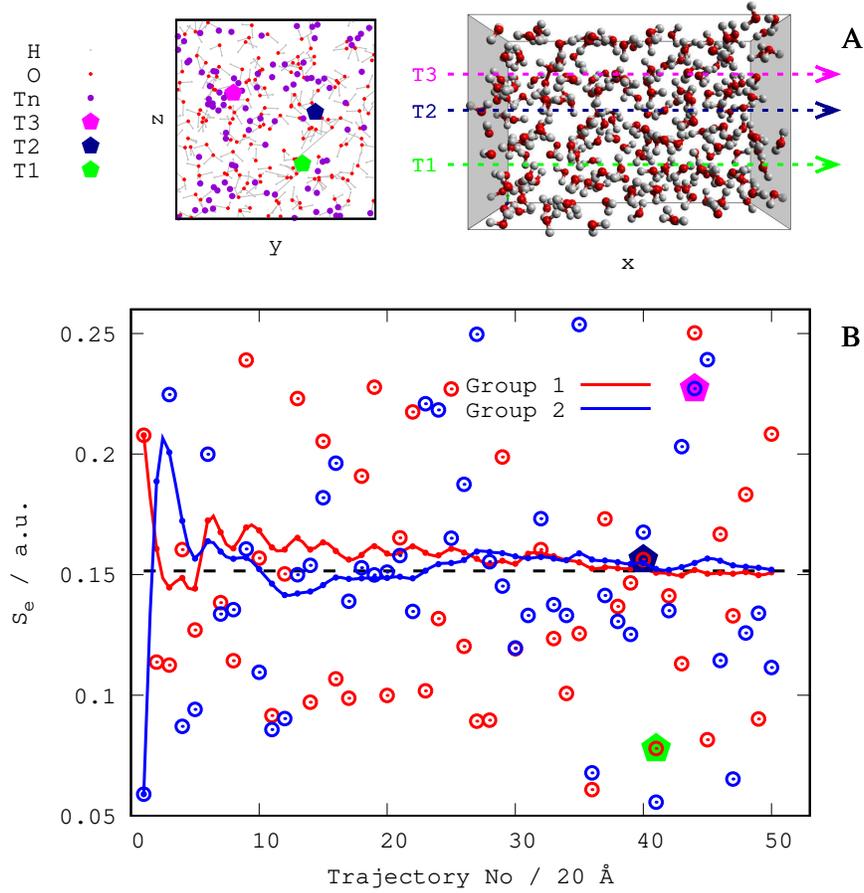}
\caption{(color online) A: Geometric configuration of a liquid water sample in a 24$\times$16$\times$16 \AA{}$^3$ box. The left panel shows a projection onto the $yz$-plane, while the right panel is a 3-dimensional representation of the box. In the rt-TDDFT simulations for the calculation of $S_e$, the proton moves along the $x$-direction at constant velocity. B: Scattered circles represent the electronic stopping power at the Bragg peak for protons in liquid water $S_e(v_p=1.72$ a.u.), calculated via rt-TDDFT for 100 short (20 \AA) trajectories along the $x$ direction, starting from random locations in the $yz$-plane. Solid lines with points show the running average of $S_e$ for two independent groups of 50 trajectories each. Note that they converge to the same value, but this requires more than 50 trajectories. Three typical trajectories with small, moderate and large $S_e$ are identified as T1, T2 and T3 in A, and as solid symbols in B.}
\label{f1}
\end{figure}

We first explored the correlation between electronic stopping power and the geometric characteristics of the trajectory by computing the stopping power for protons in liquid water for 100 trajectories at the Bragg peak energy. According to SRIM tables, the latter is located at $v_p=1.72$ a.u, corresponding to a kinetic energy of 74 keV/nucleon \cite{Ziegler2010}. The trajectories were all 20 \AA{}-long along the $x$-direction, with the initial position of the proton chosen randomly, as shown in Fig.\ref{f1}(A) for three specific
trajectories labelled T1, T2, and T3.

\begin{figure}[htbp]
\centering
\noindent\includegraphics[width=12cm]{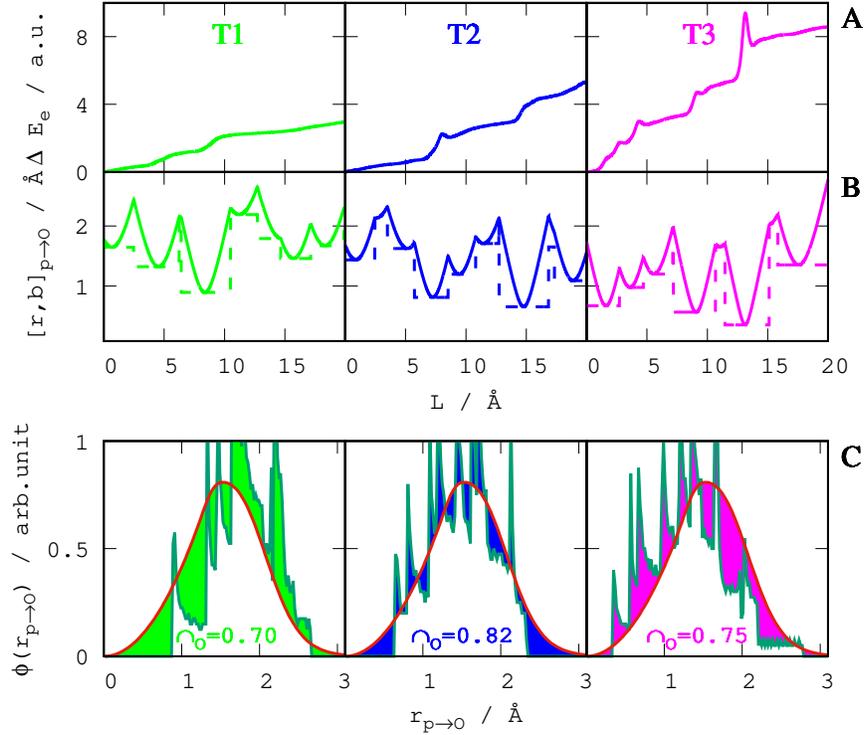}
\caption{(color online) A: Increase in electronic energy of the water sample (${\Delta}E_e$); B: distance from the projectile to the closest oxygen atom ($r_{p{\rightarrow}O}$, solid lines) and impact parameter ($b_{p{\rightarrow}O}$, dashed lines) of the proton to water oxygens; and C: probability distribution $\phi(r_{p{\rightarrow}O})$ (lines with spikes). Left, centre, and right panels correspond to the three typical trajectories T1, T2 and T3 (see Fig. \ref{f1}). The reference PDF $\phi_R(r_{p{\rightarrow}O})$ is shown in C with a red solid line, and the overlap indices, $\cap_O$, between $\phi(r_{p{\rightarrow}O})$ and $\phi_R(r_{p{\rightarrow}O})$ are reported for the three typical trajectories.}
\label{f2}
\end{figure}

The 100 values obtained for  $S_e(v_p=1.72$ a.u.$)$ are shown in Fig.\ref{f1}B with open circles. It can be seen that $S_e$ exhibits enormous fluctuations between trajectories which, for this velocity, range between 0.06 a.u. and 0.32 a.u.
Three typical situations for small, normal and large $S_e$ are shown in Fig. \ref{f1} and Fig. \ref{f2}. The example of T1 (green, trajectory 41) corresponds to a typical channeling trajectory. The proton always passes far from the nuclei in the target, with an impact parameter $b_m > 0.85$ \AA{}. As no close impact with any of the nuclei take place, the projectile travels through a region of relatively low electronic density, so that the energy of the electronic subsystem changes gently and smoothly. As a result, $S_e$ is quite small at 0.07 a.u. On the contrary, for trajectory T3 (magenta, trajectory 43) the impact parameter $b_m<0.45$ \AA{}. In particular, there is a very close collision that might involve the excitation of inner shell electrons of the oxygen atom of a specific water molecule and also other smaller jumps \cite{Yao}. Hence, $\Delta E_e$ increases much faster and experiences several jumps, leading to a much larger value $S_e=0.23$ a.u. The specific electronic excitation dynamics can be retrieved by time-dependent wave-function analysis \cite{zeb2012}, but this is outside the scope of the present work.
The case of T2 (blue) is neither a channelling trajectory nor experiences close impact events. Here $S_e=0.15$ is close to the average value obtained for the 100 random trajectories (dashed lines in Fig. \ref{f1}).

To study the convergence of the running average of $S_e$ when considering random trajectories, these 100 were divided into two groups of 50, chosen independently and randomly. It can be seen that, even with 50 trajectories, the running average is still varying. The accumulated length of the 50 trajectories is similar to that reported by Yao et al \cite{Yao}, who run 64 slightly shorter trajectories. This implies that the straightforward approach of computing $S_e$ by averaging random trajectories requires a large number of them, and is hence very demanding computationally.

The electronic stopping power in a uniform electron gas is proportional to $v_{p}$ and ${\rho_e}^{1/3}$, with $\rho_e$ the electronic density \cite{Sigmund2014,Correa2018}. The situation is more complex in real materials. In the vicinity of the projectile, the electronic density and the Kohn-Sham energy levels are significantly perturbed by the passage of the ion through the target, especially for heavy ions \cite{Lim2016}. As a result, for short, nanometer-sized trajectories, $S_e$ depends quite heavily on the specific trajectory. 
Experimentally, we have to consider many projectiles travelling much longer through the target ($\mu m$ to $mm$), thus sampling much more thoroughly the local electronic structure. Therefore, the accuracy of $S_e$ calculated by rt-TDDFT simulations can only be ensured by selecting a set of short trajectories that reproduces, to high accuracy, the sampling of the local electronic structure realized in experiment. This could be achieved either by running sufficiently many unbiased trajectories or, as we propose in this paper, by a clever choice of representative ones.

The distribution of electronic states in a material, which can be characterized by the local density of states \cite{race2013}, depends on chemical composition and bonding. It can be quite complex, but there are some general rules of thumb. For example, core electrons are tightly bound to the nuclei and hence located close to them. A close impact with core electrons will give rise to a strong retarding force on the projectile. Valence electrons have a higher probability of being located farther away from the nuclei and, hence, can be excited more easily exercising a smaller retarding force on the projectile. These qualitative arguments suggest the existence of a correlation between the local electronic structure experienced by the projectile and the geometric arrangement of the atoms of the target material relative to the projectile. 

This correlation can be employed to replace the customary on-the-fly random sampling of the local electronic structure during rt-TDDFT simulations with a trajectory pre-sampling selection tool based on a geometric criterion, to be run before carrying out any rt-TDDFT simulations. An intuitive example is the case of a monoatomic disordered system, e.g. liquid Ar, in which the electronic density is practically spherically distributed around the atoms. If we think the electronic density as a superposition of contributions from all atoms in the target, the most prominent contribution to the retarding force will arise from the atom that is closest to the projectile at any given time. Therefore the local electronic structure, and hence the electronic stopping along the projectile's trajectory, can be regarded as functions of the distance between the projectile and the closest target atom ($r_m$). Under the above assumptions, an optimal trajectory or set of trajectories, leading to an accurate determination of the electronic stopping power, should closely reproduce  a {\it reference} distance probability distribution function (PDF), $\phi_R(r_m)$, obtained by sampling $r_m$ over a rectilinear trajectory of a length that is representative of experiment.

The situation becomes a bit more complicated when instead of a monoatomic system we consider a molecular liquid like water. According to DFT calculations, in a water molecule the average electron density around the oxygen atom is more than ten times that around hydrogen \cite{Martin2005}. The symmetry of the electronic states around the water oxygen is not modified significantly by the intra-molecular O-H bonds and even less by inter-molecular hydrogen bonds. It is then reasonable to focus mainly on the electronic stopping due to oxygen atoms following the same procedure of monoatomic targets. One could then explore whether introducing the distance distribution to hydrogen atoms can improve convergence further or not.

The first step is to calculate the reference PDF, in this case for the proton-oxygen distance, $\phi_R(r_{p{\rightarrow}O})$. This was done by considering a straight trajectory in a direction incommensurate with the  simulation box that was used for the rt-TDDFT simulations. When this trajectory exits the box, it re-enters via PBC but without repeating the same path. In this way we ran the equivalent of a very long trajectory, 500 $\mu$m-long. We argue that this kind of trajectory is representative of experiment. This trajectory was then used to build the reference PDF as a histogram of the distance from the proton to the closest oxygen. The PDF arising from this trajectory is shown as a solid red line in Fig. \ref{f2}C and in the inset to Fig. \ref{f4}. $\phi_R(r_{p{\rightarrow}O})$ is an asymmetric unimodal distribution with the peak at $r_{p{\rightarrow}O}=1.62$ \AA{}. 

Once $\phi_R(r_{p{\rightarrow}O})$ is known, the goal is to reproduce it as closely as possible using a small set of short trajectories (fully contained in the simulation box), the fewer the better, as this reduces the computational cost. The electronic stopping calculated via rt-TDDFT by averaging this small set of trajectories should quickly converge to experimental observations.

To evaluate the similarity between $\phi(r_{p{\rightarrow}O})$ of an arbitrary trajectory and the reference PDF $\phi_R(r_{p{\rightarrow}O})$, we define the overlap index as: 
\begin{equation}
\cap_O=
1-0.5\times\int\,\left|\phi(r_{p{\rightarrow}O})-\phi_R(r_{p{\rightarrow}O})\right| \, dr_{p}~.
\label{eq3}
\end{equation}  
This quantity corresponds to the overlap in area under these two PDF curves. 
As the length ($L$) of the trajectory increases, the overlap index approaches 1:
\begin{equation}
\lim_{L\to\infty} \cap_O=1. 
\label{eq4}
\end{equation}

The PDF $\phi(r_{p{\rightarrow}O})$ and the overlap index of the three typical short trajectories T1, T2 and T3 are shown in Fig. \ref{f2}C. The bias of the sampling of $r_{p{\rightarrow}O}$ along them is indicated by a color-filled area, representing the difference between the two distributions. 
For T2, $\phi(r_{p{\rightarrow}O})$ generates quite a balanced sampling with a large overlap index of 0.82. As a result $S_e(\textrm{T2})$ is much closer to the average value than for the other two trajectories (see Fig. \ref{f1}B).
For T1, the $\phi(r_{p{\rightarrow}O})$ curve is skewed to right, i.e. to larger proton-oxygen distances, in comparison to $\phi_R(r_{p{\rightarrow}O})$. The proton travels mainly through the inter-molecular space, where the electronic density is relatively low. 
Therefore, the retarding force on the proton is small.
On the contrary, for T3, $\phi(r_{p{\rightarrow}O})$ is skewed to the left, i.e. to shorter proton-oxygen distances, corresponding to higher electronic density regions.
Hence, $S_e(\textrm{T3})$ is larger than the average value.

\begin{figure}[htbp]
\centering
\noindent\includegraphics[width=12cm]{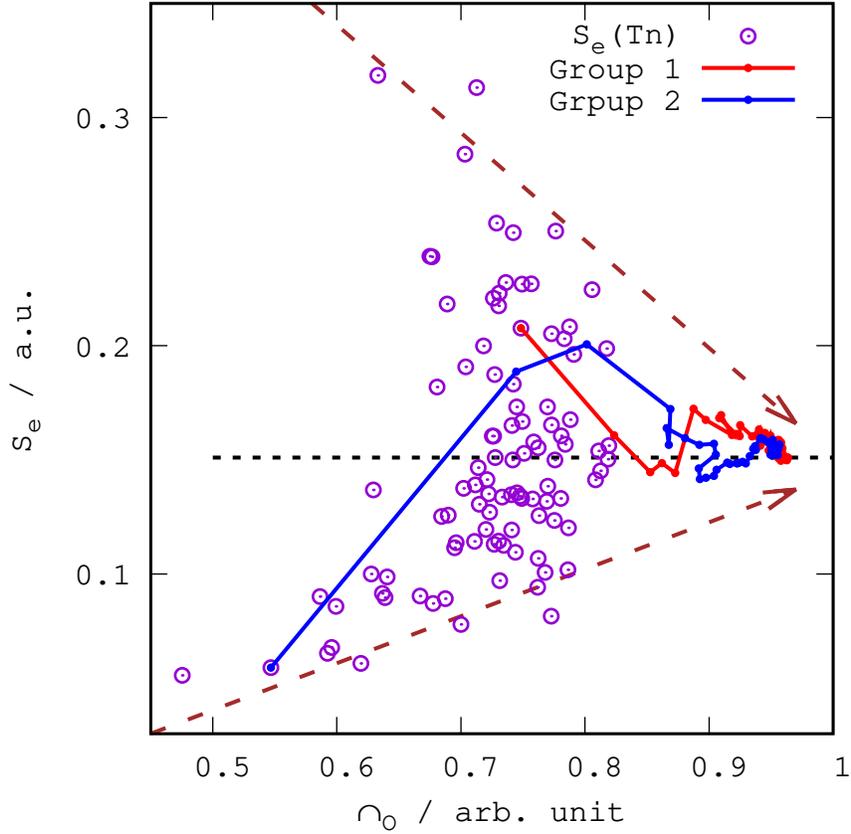}
\caption{(color online) Stopping power as a function of the overlap index $\cap_O$. The open circles correspond to 100 randomly chosen 20 \AA{}-long trajectories. The running averages of the two groups of 50 trajectories are shown with lines and points. The average of all 100 trajectories is indicated with a black horizontal dashed line. The brown dashed arrows are guides to the eye, to illustrate how $S_e$ converges with increasing $\cap_O$. }
\label{f3}
\end{figure}

In Fig. \ref{f3} we plot (open circles) the electronic stopping values $S_e(v_p=1.72 \mathrm{~a.u.})$, i.e. at the Bragg peak, for the 100 independent trajectories above, as a function of the overlap index $\cap_O$. 
At a first glance, the overlap index $\cap_O$ of individual trajectories does not exhibit a clear correlation with the calculated stopping power. In fact, for relatively small overlaps there is a substantial spread of stopping values. This simply indicates that, for short trajectories, the sampling is inadequate; achieving the correct  balance between close impact events and smooth energy transfer requires longer trajectories. However, it can be clearly observed that, as $\cap_O$ increases,the calculated values of  tend to group increasingly closer to the average value for the 100 data point (horizontal dashed line). When $\cap_O\to 1$, the value of $S_e$ obtained from rt-TDDFT calculations converges to the expected ensemble average, which should be comparable to  observation. For the present simulation box, the overlap index of short individual trajectories (open circles) does not exceed the value $\cap_O=0.85$. Therefore, more than one trajectory is required to obtain an accurate estimate of $S_e$.

The two independent running averages of $S_e$ obtained with up to 50 trajectories each, shown in Fig.\ref{f1}B, are now plotted (solid blue and red lines with points) as a function of the overlap index of the accumulated PDF. The latter is built by putting together several independent short trajectories, in random order. The zigzags in these two lines reveal the poor efficiency achieved by calculating the average $S_e$ using accumulated randomly ordered trajectories. A possible strategy to improve this situation is to replace the random order with one in which the overlap indices are taken into account.

\subsection{Geometry pre-sampling of short proton trajectories in liquid water}

In this work, the geometry pre-sampling of a set of short trajectories for protons in a liquid water sample (see Fig.\ref{f1}A) was carried according to the following algorithm.

\begin{enumerate}
    \item A number of candidate short trajectories with the proton running parallel to the $x$ axis is selected. The starting points are chosen regularly distributed over the $yz$ plane. There is no loss of generality in doing this, due to the disordered nature of the system.
    
    \item For each candidate trajectory, the overlap index $\cap_O$ is calculated.
    
    \item The trajectory with the largest $\cap_O$ is selected first. The second trajectory is the one that contributes mostly to increase the $\cap_O$ of the accumulated trajectory, i.e. by extending the first selected trajectory. This procedure is continued with other trajectories until the accumulated overlap index $\cap_O$ of all the selected short trajectories is very close to one. Notice that this procedure is different from accumulating trajectories according to their individual $\cap_O$.
\end{enumerate}
 
In order to eliminate the initial transient in rt-TDDFT simulations, and the effect of mirror excitations at the beginning and the end of the box in the $x$-direction due to periodic boundary conditions, we restrict the portion of trajectory used to compute $S_e$. For our box of length 24 \AA{}, we use, for the calculation of $\phi(r_{p{\rightarrow}O})$, $\cap_O$, and  $S_e$, the segment of length ${\Delta}L=20$ \AA{} going from $x=2$ \AA{} to $x=22$ \AA{}. In addition, violent collisions in the initial and final 2 \AA{} of each trajectory are avoided by discarding any trajectories for which 
the impact parameter $b<0.8$ \AA{}.
This empirical criterion helps excluding trajectories started and/or ended with a large variation in $E_p$ that would influence in an uncontrolled way the calculated value of $S_e$, as shown by the peaks in Fig. \ref{f2}A. This is done for both the calculations with randomly placed trajectories shown in Fig. \ref{f1}, and the geometric pre-sampling for efficient rt-TDDFT calculations. The basic algorithm and flowchart of the trajectory sampling scheme are given in Fig. S1 of the supplemental material. 

\begin{figure}[htbp!]
\centering
\noindent\includegraphics[width=12cm]{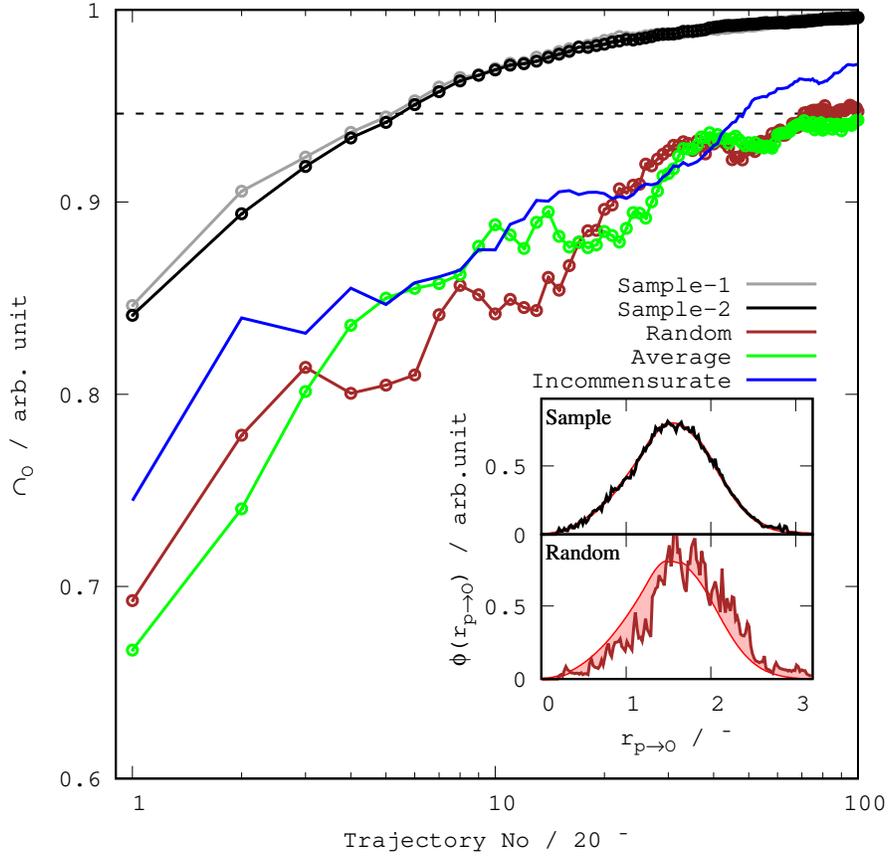}
\caption{(color online) Overlap index $\cap_O$ for increasingly large sets of 20 \AA{}-long trajectories generated by different methods: geometry pre-sampling (grey for Conf-1, and black for Conf-2), randomly (brown) and uniformly (green) distributed starting points in the $yz$ plane, and a continuous trajectory incommensurate with the main axes of the simulation box (blue). The dashed line corresponds to $\cap_O=0.95$. The inset shows the PDFs using 10 trajectories for geometry pre-sampling (upper panel) and random (lower panel), compared to the reference distribution (red solid line). Comparison of the shaded areas indicate that the former is much closer to the reference than the latter.}
\label{f4}
\end{figure}

The evolution of $\cap_O$ with the number of short trajectories generated by various methods is shown in Fig. \ref{f4}. We consider first a set of random (brown) and a set of uniformly distributed (green) short trajectories. We can see that for both these cases overlap index increases relatively slowly with the number of trajectories. With 10 trajectories, the overlap index is $\cap_O\approx 0.85$, a value that can be achieved with a single trajectory using geometry pre-sampling. With 10 pre-sampled trajectories, the overlap index can be increased to $\cap_O\approx 0.97$, so that $\phi(r_{p{\rightarrow}O})$ is almost indistinguishable from the reference distribution, which represents the  real observation.

Correa {\it et al.} have suggested that, in a periodic crystal, a rectilinear trajectory incommensurate with the lattice will explore all the possible distances (impact parameters) with geometrically correct weights, but only when the trajectory is sufficiently long \cite{Correa2018,Schleife2015}. 
To assess the effectiveness of this scheme in liquid water as a representative disordered material, we calculated the overlap index $\cap_O$ for a trajectory that is not parallel to any main axis of the simulation box. This is reported in Fig. \ref{f4} (blue solid line). It can be seen that the evolution of $\cap_O$ with this method is similar to that for random and uniformly distributed sets of trajectories. 


\section{\label{sec3} Results and discussion}
Using the pre-sampled short trajectories we calculated the electronic stopping power $S_e(v_p)$ for protons in liquid water as a function of velocity, in the range $v_p=0.2 - 8$ a.u., which includes the Bragg peak. This was done by means of rt-TDDFT simulations at the theory level and basis set described in Section \ref{sec2.1}. 
\begin{figure}[htbp]
\centering
\noindent\includegraphics[width=12cm]{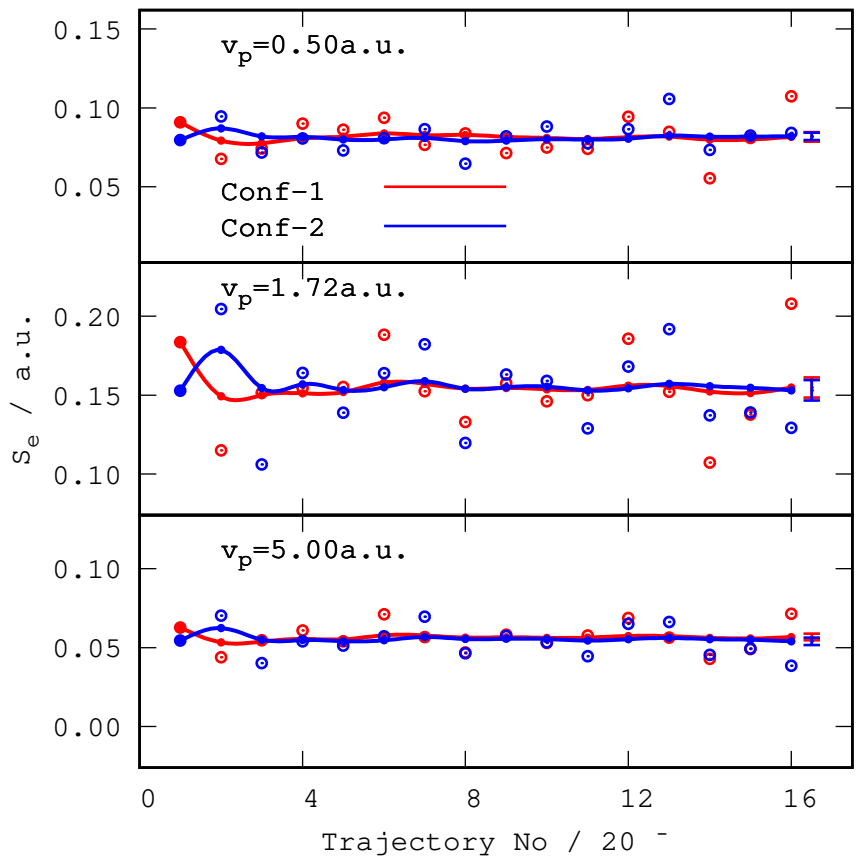}
\caption{(color online) Stopping power for protons in liquid water calculated with pre-sampled short trajectories. The running averages of $S_e(v_p)$ for three typical velocities, 0.5, 1.72 and 5 a.u., are shown with solid lines. The standard deviations are shown as a vertical tick to the right of each line. Clearly, the largest uncertainty corresponds to the Bragg peak velocity (1.72 a.u.)
The individual $S_e(v_p)$ values for each selected short trajectory are indicated with open circles. Results for two different water configurations (Conf-1 and Conf-2) are displayed in red and blue, respectively.}
\label{f5}
\end{figure}
Fig. \ref{f5} shows $S_e(v_p)$ for protons at three different velocities: 0.5 a.u., 1.72 a.u. and 5 a.u., i.e. below, at, and above the Bragg peak, for 16 pre-sampled short trajectories for the two different liquid water samples, Conf-1 (red circles) and Conf-2 (blue circles) introduced in Section \ref{sec2.1}. The initial proton coordinates of the selected trajectories are given in Table S2 of the supplemental material. The running averages of $S_e$ are shown with (red and blue) solid lines and points. For the sake of clarity, the standard deviations from the 16 $S_e$ data points are shown only at the end of the running average lines. 

The first observation is that the results for the two independent configurations present the same accuracy. 
Although the values of $S_e$ for the individual short trajectories are quite different for Conf-1 and Conf-2, the running averages converge to the same value. This means that, for disordered systems like liquid water, the pre-sampling scheme is independent of the snapshot selected for the calculation of $S_e$. In addition, as shown in Fig. \ref{f4}, after the first three trajectories the difference in the overlap index $\cap_O$ between the two configurations is extremely small. Hence, the efficiency of the pre-sampled scheme is also insensitive to the configuration of the target. Any configuration extracted from a stable adiabatic {\it ab inito} MD trajectory can be used to calculate $S_e$.

The second and most important observation is that using the pre-sampled trajectories the calculated $S_e$ converges much faster than with any of the other three schemes, namely random distribution, uniform distribution, and long incommensurate trajectory. For  $v_p=1.72$ a.u., which corresponds to the Bragg peak in SRIM data tables, the running average of $S_e$ converges to 0.154 a.u. using only eight pre-sampled trajectories, corresponding to an accumulated length of 160 \AA{}. After that, the running average remains stable within $\pm 0.003$ a.u., i.e. < 2\%. The difference between the average computed with eight pre-sampled trajectories and the average obtained using 100 random trajectories (0.152 a.u.) is only 0.002 a.u., i.e. less than 1.5\%, at about 10\% of the computational effort. The cases $v_p=0.5$ a.u. and $v_p=5$ a.u. correspond to low and high velocities, respectively. Here, only four pre-sampled trajectories, with an accumulated length of 80 \AA{}, are needed for the running average of $S_e$ to converge to a similar accuracy. This indicates that $S_e$ becomes less sensitive to the trajectory
when moving away from the Bragg peak, thus demonstrating the influence of the local electronic states on $S_e$. Therefore, the accuracy of the stopping curve as a function of velocity is dominated by the accuracy at the Bragg peak.
To the best of our knowledge this is the first time the convergence of rt-TDDFT $S_e$ calculations is discussed for velocities around the Bragg peak. Existing convergence benchmarks for protons in liquid water have been carried out for high proton velocities ($v_p=8.$ a.u.) with uniformly distributed short trajectories \cite{Yao}. For protons in crystalline aluminium, benchmarks were performed for low and high proton velocities ($v_p=0.1$ and $v_p>6$ a.u.) with long incommensurate trajectories \cite{Schleife2015}. 
 
\begin{figure}[htbp]
\centering
\noindent\includegraphics[width=12cm]{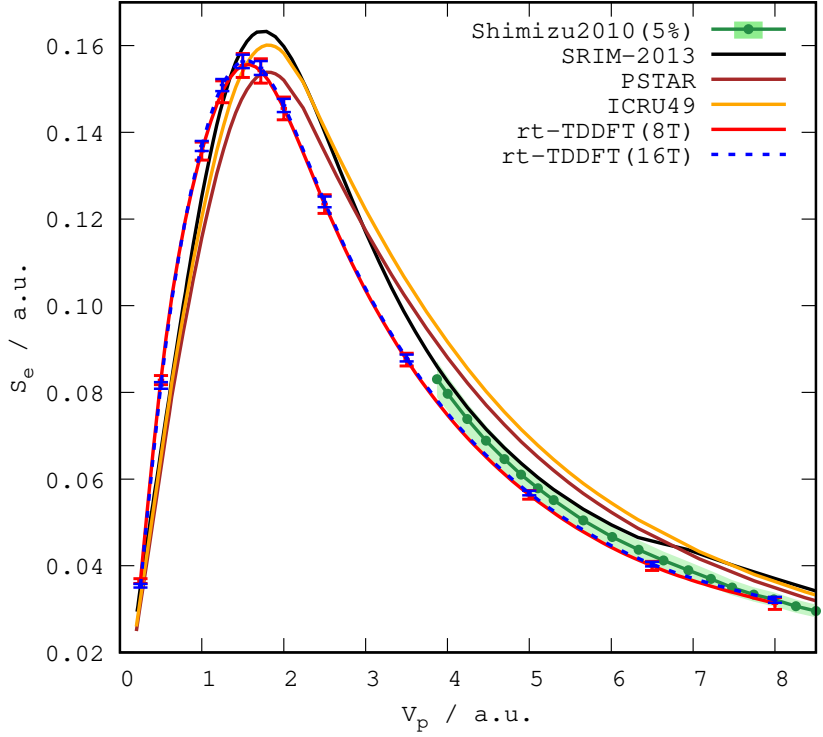}
\caption{(color online) Stopping power curve $S_e(v_p)$ calculated by rt-TDDFT using 8 (red) and 16 (blue dash) pre-sampled short trajectories, along with experimental data \cite{Shimizu2010} (Shimizu-2010, green symbols, the confidence interval ($\pm$5\%) is shown as a light green shaded area), and empirical data taken from SRIM-2013 \cite{Ziegler2010} (black line), PSTAR \cite{pstar} (brown line), and ICRU49 \cite{ICRU49} (orange line) tables. Error bars for the rt-TDDFT calculations correspond to the standard deviation using 8 or 16 trajectories.}
\label{f6}
\end{figure}
 
Fig. \ref{f6} shows the entire stopping curve $S_e(v_p)$ for protons  in liquid water, for  12 different velocities spanning the range from 0.2 au to 8.0 au,  using the first eight (dashed red line) and 16 (solid blue line) pre-sampled short trajectories based on Conf-1. The data are provided in Table S3. The two curves overlap with each other, indicating that eight trajectories are sufficient at all velocities. The computational resources used are less than 20\% those required with uniform \cite{Yao} or randomly trajectories. This demonstrates that the geometry pre-sampling scheme proposed here, based on maximizing the overlap index $\cap_O$, is an efficient, accurate, and robust for the rt-TDDFT calculation of $S_e(v_p)$. 

In Fig. \ref{f6} we also report the experimental data by Shimizu {\it et al.} for protons in liquid water for $v_p> 3.5$ a.u.  \cite{Shimizu2010} and the empirical electronic stopping data according to SRIM-2013 \cite{Ziegler2010}, PSTAR \cite{pstar} and ICRU49 \cite{ICRU49} tables. For $v_p > 5$ a.u. the rt-TDDFT results are all located within the confidence interval ($\pm$5\%) of experimental observations (green dots and light green shaded area in Fig. \ref{f6}). The difference between them decrease to less than 2\% for $v_p=8.0$  a.u. 
At low velocities, around $v_p=0.25$ a.u., rt-TDDFT results are very close to all, SRIM, PSTAR and ICRU49 values. However, in the Bragg peak region there are some interesting discrepancies. While for SRIM and PSTAR the peak is located at similar velocities, the stopping values at the peak are different: 0.162 a.u. (SRIM) and 0.151 a.u. (PSTAR). The value $S_e=0.154$ a.u. obtained by rt-TDDFT is bracketed between these two values, and is very close to that of ICRU49 (0.158 a.u.). However, the maximum of rt-TDDFT calculations is located at $v_p=1.55$ a.u., which is lower than both $v_p=1.72$ a.u. of SRIM and PSTAR, and $v_p=1.80$ a.u. of ICRU49, i.e. there is a red shift of around 0.2 a.u. with respect to these empirical curves. A similar red-shift has also been observed in rt-TDDFT calculations for protons in lithium by Maliyov {\it et al.} using Gaussian basis sets\cite{Maliyov2018}, and for protons in aluminum by Correa {\it et al.} using plane wave basis sets \citep{Schleife2015}.
All these results seem to indicate that the red-shift is not related to a technical problem of the simulations. Nevertheless,
further analysis of the influence of basis set type and size is underway, and will be presented elsewhere.  Interestingly, a red-shift of the electronic stopping power curve for protons in water vapor compared to SRIM-2006 table is also reported in calculations using the quantum-mechanically based Monte Carlo track-structure code TILDA-V\cite{Alcocer-Avila2019}. 
An additional advantage of the pre-sampling method, being dependent only on geometric parameters, is that the set of selected short trajectories can be used for the calculation of stopping power for different projectiles like $\alpha$-particles and Carbon ions. 

It could be argued that the consideration of a long, constant-$v$, rectilinear trajectory for generating the reference distance probability would give rise to a red shift of the Bragg peak, since the lower the velocity the more likely the deflection of the trajectory when passing close to target nuclei. Hence, the calculated stopping power with that reference for low velocity would effectively sample shorter distances than expected in experiment. It is easy to see, however, that such deviation is expected
to be negligible at Bragg-peak velocities. Deflections significantly affecting the minimum distance reached by the projectile to any target nucleus would require Coulombic repulsion between the proton and the target Oxygen of the same scale as the kinetic energy of the projectile, $E_k \sim Z/d_{min}$, which for energies around the Bragg peak implies distances in the scale of the m\AA. The reference PDF (see e.g. Fig.~\ref{f2}C or inset to Fig.~\ref{f4}) is extremely low on that scale and, therefore, the suspected overestimation of the electronic stopping due to this effect would be undetectable on the scale of the results in Fig.~\ref{f6}. This aspect could, however, affect the results in the low-velocity region, and it is an interesting avenue for further investigation.

In a strict sense, the condition $\cap_O=1$ is not sufficient 
to guarantee an ergodic exploration of the electronic states experienced by the projectile. The success of this simple geometry pre-sampling method in the case of a water target is partially due to the fact that the average electronic density around oxygen is much higher than around hydrogen. We have carried out additional tests using a combination of overlap indices for oxygen ($\cap_O$) and hydrogen ($\cap_H$), namely $\cap_X=0.7\,\cap_O+\,0.3\,\cap_H$, and observed that $S_e$ converges very quickly to the same value obtained using $\cap_O$, albeit the pre-sampled trajectories are not necessarily the same. For disordered or non-periodic systems with many kinds of atoms and complex chemical bonds, the one-dimensional (radial) geometry pre-sampling criterion used in this work should be extended to a multi-dimensional approach, as valence electrons are often located along chemical bonds instead of around atoms. Water is a particular case because the electronic density is mostly concentrated around the oxygen atoms, but this is not the case in DNA or living cells. A proper assessment of energy deposition in the latter is crucial within the context of space travel and radiotherapies.
To match the stopping power calculated via rt-TDDFT with observations, and thus render the calculations useful for applications, it is important to consider the chemical and geometric structure as well as the concentration of biomolecules like DNA and proteins, in vivo.

It should be mentioned that, although the impact parameter $b$ is widely used in studying two-body collision events in radiation physics, it is probably not a suitable parameter to describe disordered extended systems, as a given $b$ does not correspond to a unique value of the closest distance $r_{p\to O}$ (see dashed lines in Fig. \ref{f2}B). This is not necessarily the case of crystalline systems. Maliyov {\it et al.} developed an averaging scheme that considers the dependence of the stopping power on impact parameter, including its angular dependence, $S_e(v,\mathbf{b})$, and successfully applied it to metallic systems, notably Li and Al \cite{Maliyov2018,Maliyov2020}. Since no error bars are reported in these works, it is difficult to assess the statistical accuracy of this attractive method.

\section{\label{sec:4}Conclusions and perspectives}

In this paper we proposed a new method to compute accurately the electronic stopping power in disordered materials at a reduced computational cost. The approach is based on selecting an optimal set of short, nanometer-size trajectories beforehand using a purely geometric criterion, thus limiting the number of costly rt-TDDFT simulations. The geometric criterion consists of imposing that the probability distribution function (PDF) of the projectile-nearest target atom distance obtained by accumulating the short trajectories reconstructs as closely as possible the reference PDF corresponding to a very long trajectory, representative of the experimental situation. This exploits the correlation between distance and electronic density, and the known dependence of electronic stopping on the latter. It does not, however, take into account the dependence of stopping on the local electronic structure.

In the implementation of this scheme for protons in liquid water, which is a typical disordered target and the most essential component of biological tissue, $S_e(v_p)$ converges very quickly with the number of short trajectories, requiring at most eight 20 \AA{}-long trajectories. The most demanding situation corresponds to the Bragg peak, where the dependence of $S_e(v_p)$ on the local electronic structure along the trajectory is most prominent. Other schemes frequently used, like a random or a uniform distribution of trajectories, require approximately ten times more samples to achieve the same level of accuracy, thus making them  too demanding for most present applications. The stopping curve calculated via rt-TDDFT is comparable with existing experimental data at high velocity, and with SRIM and PSTAR tables at low velocity. The main difference between our rt-TDDFT calculations and existing empirical stopping curves is located in the Bragg peak region. Similarly to calculations by other authors in different systems, the position of the rt-TDDFT peak is red-shifted, while the maximum value of $S_e$ lies between SRIM and PSTAR, and close to ICRU49.

The high accuracy and efficiency of the present geometry pre-sampling scheme could be helpful in promoting the application of rt-TDDFT simulations to the computation of stopping power in a variety of disordered systems, generating results that are comparable to observations at a  moderate computational cost. The extension of this methodology to crystalline systems is straightforward, and is currently being used to compute electronic stopping in water ice. Moreover, having this level of accuracy and control over error bars could be useful to assess and possibly re-calibrate existing stopping tables for compounds that are presently based on the straight Bragg rule \cite{Thwaites1983,Quashie,Ziegler2010}, thus avoiding {\it ad hoc} combinations and scaling factors.  Interestingly, recent work has shown that the straight Bragg rule works poorly for oxides (water is considered an oxide) at low velocities when comparing to experiment \cite{Sigmund2018}. More elaborate models taking into account the role of the oxygen $2p$ electrons \cite{Roth2017} do improve over Bragg-based approaches, in a system-dependent way \cite{Sigmund2018}. In any case, the latter are not the models used to calculate SRIM and PSTAR tables.


This approach requires extensions if it is to be applied to more complex systems, in which the electronic density does not correlate directly with the distance between the projectile and the atoms in the sample, or in which there is an important dependence of the local electronic structure along the trajectory. This may happen if there are more than one dominant species as in biomolecular systems, where C, N, and O are generally present in comparable amounts, or in binary semiconductors like GaAs. 

Another situation of interest arises when considering mixtures or solutions composed of two or more parts, like a DNA fragment or even a whole nucleosome in the physiological environment. For these cases, an accurate description of stopping power would be highly desirable, as present-day Monte Carlo track structure codes only calculate the transport through water. If one can produce {\it ab initio} data for the various individual parts, then if would be possible to compute the stopping power  of the whole, and hence the energy deposited along a track, by adding the contributions of the various components, using a simple approach along the lines of Bragg's rule or a more advanced model for compounds like PASS \cite{Sigmund2002,Sigmund2018}.

\begin{acknowledgments}
This work has received funding from the Research Executive Agency under the EU's Horizon 2020 Research and Innovation program ESC2RAD (grant ID 776410). We are grateful for computational support from the UK national high performance computing service, ARCHER, for which access was obtained via the UKCP consortium and funded by EPSRC grant ref EP/P022561/1. \\

\noindent \textbf{DATA AVAILABILITY}

The data that support the findings of this study are available from the corresponding author upon reasonable request.
 
\end{acknowledgments}

\bibliography{apssamp}
\end{document}